# Hazard Analysis for Self-Adaptive Systems Using System-Theoretic Process Analysis

Simon Diemert
*Department of Computer Science*
*University of Victoria*
Victoria, Canada
sdiemert@uvic.ca

Jens H. Weber
*Department of Computer Science*
*University of Victoria*
Victoria, Canada
jens@uvic.ca

*Abstract*—Self-adaptive systems are able to change their behaviour at run-time in response to changes. Self-adaptation is an important strategy for managing uncertainty that is present during the design of modern systems, such as autonomous vehicles. However, assuring the safety of self-adaptive systems remains a challenge, particularly when the adaptations have an impact on safety-critical functions. The field of safety engineering has established practices for analyzing the safety of systems. System Theoretic Process and Analysis (STPA) is a hazard analysis method that is well-suited for self-adaptive systems. This paper describes a design-time extension of STPA for self-adaptive systems. Then, it derives a reference model and analysis obligations to support the STPA activities. The method is applied to three self-adaptive systems described in the literature. The results demonstrate that STPA, when used in the manner described, is an applicable hazard analysis method for safety-critical self-adaptive systems.

## I. INTRODUCTION

Modern systems are increasingly required to manage uncertainty arising from variable and evolving operating conditions. Adaptation is an important mechanism for addressing this uncertainty. A system that can change its behaviour or structure in response to changes in itself or the environment is called a "self-adaptive system" (SAS) [1]. Adaptation has been used in a wide range of applications, including web service management [2]–[4], wireless sensor networks [1], [5]–[7], and robotics and autonomous systems [8]–[13].

Many applications areas exist where a SAS might contribute to an accident, i.e., a loss event resulting in harm to persons, the environment, or property [14]. For instance, a self-adaptive function in an autonomous ground vehicle might incorrectly adapt the characteristics of the electronic brake controller such that braking is less effective than appropriate for the current conditions. SASs that perform safety-critical functions are called "safety-critical self-adaptive systems" [15]. There is a need to develop methods to assure SASs used in high-risk application areas [16]. Indeed, assurance of SASs, for safety and other quality attributes such as reliability or efficiency, has been the subject of many research initiatives [13], [17]–[19].

Established engineering practices exist to assure conventional (non-adaptive) systems. These practices are described in standards that prescribe an engineering process model to build confidence that a system is acceptably safe [20]–[23]. *Hazard analysis* is an essential activity that includes identifying hazards, estimating their criticality, and analyzing designs to determine hazard causes. Methods such as Fault Tree Analysis (FTA), Failure Modes and Effects Analysis (FMEA), Hazardous Operability Study (HAZOP) are used for this purpose [20], [23], [24]. Once hazard causes are identified, standards require that safety requirements (i.e., 'mitigations') are implemented to mitigate risk.

State of the art hazard analysis methods for SAS either involve design-time only methods [25]–[28] or hybrid (design- and run-time) methods [17], [29]–[31]. The hybrid methods identify hazard causes using conventional design-time analysis (e.g., FMEA) and then use various approaches to assess whether the system is vulnerable to those causes at run-time. Existing approaches are limited in three ways. First, they use hazard analysis methods that employ a limited notion of accident causality [14]. Specifically, they assume that hazards occur due to technical component failures (e.g., stuck valves, memory corruptions, short circuits etc.). This assumption was appropriate for simple systems when those methods were first created; however, it is not applicable to more complex systems, such as SASs, where hazards might also occur as a result of interactions between components [14], [32], [33]. Second, existing methods focus on hazard causes arising in the managed system (which performs the system functions); they do not consider the possibility that the managing system (which performs adaptations) might contribute to the occurrence of a hazard. However, since managing systems are typically complex pieces of software, it is possible that they will also experience failures, which might impact safety-related functions of the managed system [16]. Third, while existing methods use the outputs of conventional analysis methods (like FMEAs), they do not explicitly consider SAS-specific aspects as part of the design-time activities that generate the hazard causes. It is common to support conventional hazard analyses with system- or technology-specific reference models or guide phrases to aid experts. However, SAS-specific guidance is not available and so some hazard causes might be overlooked.

In summary, for SASs to be accepted by stakeholders and regulators for use in safety-critical applications, it is necessary to show that the entire SAS (including the managing system) has been subjected to a systematic hazard analysis that takes account of SAS-specific aspects. It follows that there is a need for hazard analysis methods that adopt a wholistic notion of accident causality that includes the possibility that the behaviour of the managed system contributes to the occurrence of a hazard.

### A. Contribution - STPA for Self-Adaptive Systems

While performing hazard analysis at run-time is an important line of inquiry [34], the work described in the present paper focuses on advancing design-time hazard analysis using a system-theoretic approach that incorporates details from SAS reference architectures.

System-Theoretic Process Analysis (STPA) is a relatively new (on the scale of safety engineering) hazard analysis method developed by Leveson [14], [32]. An essential idea of STPA is that hazards occur not only from component failures but also from nominal interactions between system components, including the possibility that a component or the environment is changed (adapted or evolves). STPA uses a system modelling approach called System-Theoretic Accident Model and Processes (STAMP) to describe relationships between system components using feedback control loops, not unlike the notion of feedback control and Monitor-Analyze-Plan-Execute with Knowledge (MAPE-K) paradigm that has been widely used for SASs [35], [36]. Given their conceptual parallels, STPA is a natural fit for SAS hazard analysis.

Even with synergy between STPA and SASs and the advantages of STPA over other hazard analysis methods [14], [37], [38], STPA is ultimately a general-purpose method. Blindly applying STPA without an understanding of how to model a SAS using STAMP or what it means for a SAS to be "safe" might result in an inadequate analysis. Therefore, it is desirable to tailor Leveson's generic STPA method to SASs, as has been done for other application areas [39]–[42].

This paper extends Leveson's original STPA to self-adaptive systems based on the type of adaptations performed by the system. The extended method is applied at design-time to systematically identify hazard causes originating from both the managed and managing systems. In addition to finding hazard causes, the method also identifies safety requirements for both managed and managing system to reduce risk. As part of the extended method, this paper derives a reference STAMP control structure for a SAS and describes analysis obligations, i.e., conditions to consider when performing STPA. The extended method is evaluated by applying it to three SASs. To our knowledge, this is the first paper to apply STPA to SASs.

## II. FOUNDATIONS

Per Weyns, a SAS is a system that complies with two principles [1]. The External Principle says that a SAS must be able to autonomously handle changes and uncertainties from the environment, within itself, and its goals. The Internal Principle says that a SAS is made of two main parts: one that interacts with the environment and another deals with adaptation concerns and uses a feedback loop to interact with the first part. The part of the SAS that interacts with the environment is called the *managed system* whereas the part of the system that deals with adaptation is called the *managing system* [36]. Adaptation goals, that describe high-level objectives to be satisfied, are provided to the managed system. The MAPE-K reference model is widely employed to realize the adaptation capabilities of the managing system [35], [36] and is the basis of the STAMP reference control structure in Section III.

### A. Safety-Critical Self-Adaptive Systems

Previous work has established definitions and concepts for safety-critical SASs that inform the current paper [15], they are briefly re-accounted here.

Hazards are a foundational concept for system safety engineering activities and are the conceptual "bridge" between the system and its operating environment; they are the top-level events from which all other safety engineering activities flow. More formally, a *hazard* is "a system state or set of conditions that, together with a particular set of worst-case environment conditions, will lead to an accident (loss)" [14]. A *safety-critical system* is one whose incorrect, inadvertent, or out-of-sequence action(s), in combination with conditions in the environment, contributes to the occurrence of (or failure to mitigate effect of) a hazard [23], [43]. A system is a *safety-critical self-adaptive system* if: i) it satisfies the External Principle, ii) satisfies the Internal Principle, iii) and the managed system is safety-critical [15].

An *adaptation option* is a candidate configuration for the managed system that the managing system generates to satisfy adaptation goals [1]. A *safe adaptation option* is an adaptation option that "when applied to the managed system, does not result in, or contribute to, the managed system reaching a hazardous state" [15]. An adaptation option can be demonstrated to be safe via design-time or run-time analysis activities such as testing, simulation, and formal verification. Typically, the results of these activities are captured in an assurance case that might be dynamically updated with new assurance evidence as the system adapts [16], [18], [44], [45].

Adaptations performed in a SAS may be classified according to the impact (assuming nominal operation of the SAS) they have on the safety-critical function(s) of the managed system [15]. The classification of adaptations is an important step in the extended STPA method described in Section III. There are four types of adaptations:

*1) Type 0 - Non-Interference:* These adaptations do not affect safety-critical functions of the managed system.

*2) Type I - Static Assurance:* These adaptations (and Type II and III below) affect safety-critical functions of the managed system. The managing system only selects adaptation options from a pre-determined set of safe options. Each option is shown at design-time to be safe regardless of the operational conditions the system might encounter.

*3) Type II - Constrained Assurance:* Like Type I, the managing system selects adaptation options from a pre-determined

set. However, their safety depends on the state of the environment or system. Prior to applying an option, the managing system updates a dynamic assurance case with evidence that the safety conditions for the option are satisfied. A special case (Type IIb) is where each successive adaptation monotonically increases the strength of assumptions (i.e., becoming more restrictive) on the state of the environment or system.

*4) Type III - Dynamic Assurance:* The set of adaptation options is not defined or cannot be shown to be safe at design-time. Adaptation options must be assessed at run-time and the results are incorporated into a dynamic safety case as part of the adaptation operation.

In previous work [15], for Type II adaptation we also required that constraints from subsequent adaptations are monotonically increasing. However, in this paper we use a more general definition that permits a system to return to a previous configuration provided the operating conditions for the configuration are satisfied.

*B. STAMP and STPA*

Leveson advocates for wholistic notion of accident causality where "accidents are seen as resulting from inadequate control or enforcement of constraints on safety-related behaviour..." [14, pp. 100]. It follows that safety is a control problem rather than a component reliability problem. From this perspective, STAMP offers a model of accident causation based on feedback and control flow. A STAMP model is a 'control structure' that describes functional (control) relationships between elements of a system.

A simple STAMP control structure is depicted in Fig. 1. Elements in a STAMP control structure are either: Processes that are subject to control, Controllers that provide control actions, Sensors that observe the state, or Actuators that carry out control actions. A Controller may also have an internal Process Model that models the state and behaviour of elements it interacts with; the Process Model is an imperfect reflection of reality and might be inaccurate or incorrect. Control structures may be nested hierarchically to reflect real-world relationships between systems. STPA is a hazard analysis method that is based on STAMP's causality model. Notably, STPA helps an analyst to consider causal factors beyond only component failure, including control conflicts and interactions, human decision making and cognition errors, social/organizational factors, systematic design defects (e.g., software bug) [14]. The overall objective of STPA is two-fold: to identify ways that hazards can occur and then to identify mitigations that will prevent hazard occurrences. STPA has four steps [46]:

*1) Identify Losses and Hazards:* Understand the "safety problem" at hand by identifying the unacceptable loss events. Then identify the hazardous conditions that, if they occur, can lead to a loss.

*2) Model Control Structure:* Using the STAMP modelling paradigm and an appropriate level of abstraction, describe the system as a control structure.

*3) Identify Unsafe Control Actions:* Using guide phrases ('provided', 'not provided', 'too late/too early', 'too long/stop

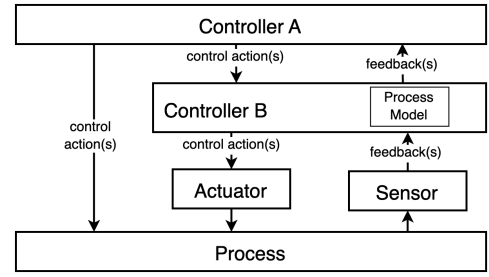

Fig. 1. Simple STPA control structure.

too soon'), each control action in the control structure is analyzed to see if it could contribute to a hazard occurrence, i.e., if it is an unsafe control action (UCA). A short standard-form sentence describing each UCA is recorded.

*4) Identify Loss/Causal Scenarios:* For each UCA, identify one or more scenarios that show how the UCA might contribute to a hazard occurrence. In this step analysts consider causal factors for the offending Controller and other elements. A reference diagram of a generic control structure annotated with suggestions for causal factors is used as a guide.

Mitigations may be identified after either Step 3 or Step 4. In some cases, a UCA can be mitigated immediately, in other cases further analysis is necessary to understand the mitigations that are required.

Several variants and extensions of STAMP and STPA have been developed: CAST is a complementary method used to determine how an accident happened, after the fact [14]; Thomas formalized some aspects of STPA to generate safety requirements [47]; Young and Leveson extended the method to include cybersecurity [39]; Stringfellow and France both extended the method to include human and organizational factors [48], [49]; Weber and Mason-Blakely extended STAMP for electronic medical records [41].

### III. STPA FOR SELF-ADAPTIVE SYSTEMS

SASs are naturally described using feedback control loops and are thus amendable to modelling using STAMP and analysis with STPA. This section tailors the existing STPA methodology to SASs by 1) extending the analysis method with additional steps to identify and classify adaptations, 2) deriving a specialized reference STAMP control structure for SASs, and 3) identifying analysis obligations that can be used to guide the analysis.

*A. Extended STPA Method for Self-Adaptive Systems*

Fig. 2 shows an overview of the extended STPA method for SASs. Analysis steps are shown in rounded rectangles with bold text and progression between analysis steps ("control flow") is depicted with solid back arrows. Steps for a conventional STPA are shown in black/white boxes and the extension for SASs is shown with shaded blue boxes/arrows. The conventional STPA analysis method was described above in Section II; each new element is discussed in detail below.

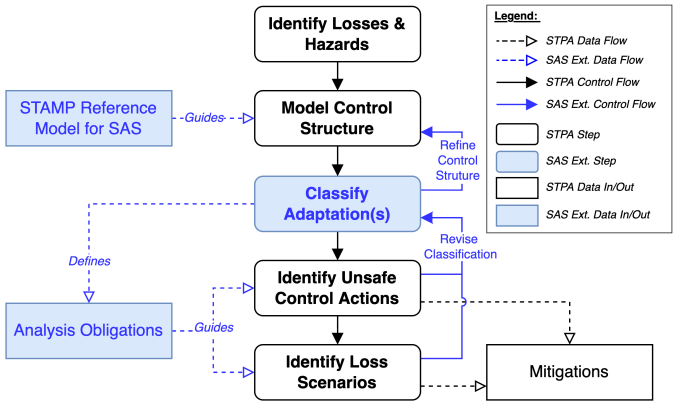

Fig. 2. Overview of extended STPA method for SAS.

*1) STAMP Reference Model:* Modelling of the STAMP control structure is guided by our STAMP Reference Model for SASs (see Section III-B). Reference models describe modelling patterns that an analyst may incorporate into modelling for a specific system. STAMP reference models for other tailored versions of STPA have been developed; for example Mason-Blakey and Weber describe a reference model that tailor's STAMP to a health system that incorporates electronic medical records [41], [42].

*2) Classify Adaptation(s) and Analysis Obligations:* After the control structure is modelled, the adaptation(s) performed by the SAS should be identified and classified. Adaptations are classified into one of four "types" (Type 0, I, II, III) using the taxonomy for safety-critical adaptations [15]. The type of adaptation is used to define analysis obligations that guide the identification of UCAs and loss scenarios.

One strength of STAMP control structures is that the level of detail in the model may be adjusted to suit the needs of the analysis: the modelled control structure should contain enough detail to allow analysts to reason about adaptations and their impact on safety. Iterative refinement may be used to increase the level of detail as needed. Additionally, when identifying UCAs or loss scenarios, the analyst might be prompted to re-consider the classification of an adaptation. Therefore, a revision feedback step is also shown in Fig. 2. Finally, though not depicted because it is part of the original STPA methodology, an analysis may opt to refine the control structure while identifying unsafe control actions or loss scenarios.

### B. STAMP Reference Model for Self-Adaptive Systems

This section uses a sequence of modelling steps to successively derive a STAMP control structure for a generic SAS. The derivation (and our rationale) are presented as an informal proof-by-construction intended to show validity of final reference model in Fig. 3. The modelling steps can also applied to a real system to arrive at the corresponding control structure.

In the system diagrams, components are shown as unshaded rectangles and are annotated with a 'badge' indicating their element type. For example, controller type elements are annotated with a "C" and probe type components are annotated with a "Pb". Shaded boxes are used to delineate between the managed and managing system.

*1) Initial Model:* The derivation begins with the managed and managing system. The managing system performs adaptation actions on the managed system and monitors values from the managed system. When deriving a model for a real system, the specific adaptation actions and monitored variables should be described in this step.

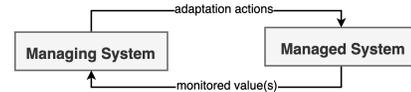

*2) Refine Managed System:* The managed system is refined to include at least one Process and optionally a Controller. Depending on the system, multiple Controllers and Processes may be modelled. Processes either represent a quantity subjected to control or might represent aspects of the environment. Multiple Sensor and Actuator components may be optionally included in the control structure for managed system. Depending on the desired level of detail, including Sensors and Actuators might not be necessary: it is possible to perform STPA without them, and they are not modelled in many published examples [14].

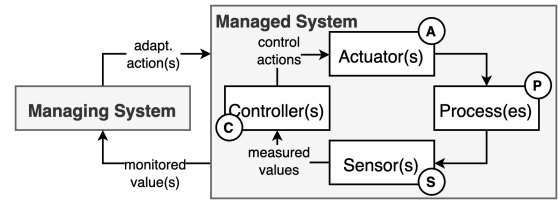

*3) Add External Control Inputs and Disturbances from Environment:* Other control inputs and external disturbances are added to the managed system. These arise from unmodelled Controllers or Processes, or the environment and might affect elements in the managed system. This step is not necessary if inputs and disturbances are not relevant for the analysis.

*4) Add Probes and Effectors:* Probes and Effectors form an interface between the managing and managed systems. Probes monitor specific aspects and provide data to the managing system. Effectors change the configuration of the managed system based on adaptation actions. At least one Probe and one Effector should be included. Each Controller, Actuator, or Sensor may be connected to a Probe or Effector. The rationale for permitting Probes and Effectors to connect to a Controller is obvious. For Sensors, since they provide measurements of the Process(es) that might be necessary for the managing system, they may be connected to a Probe. The behaviour of a Sensor might be changed (e.g., tuning its sensitivity) by the managing system and so may be connected to an Effector. For Actuators, the managing system might change their behaviour (e.g., enable/disable actuators), so they may be connected to an Effector. Additionally, the managing system might need to monitor properties (e.g., physical performance) of an Actuator, so it may be connected to a Probe. Per Weyns, Probes may also be used to directly monitor the Process(es) within the managed system or the environment [1]. Effectors may make changes to Process(es) directly but are not able to directly influence the environment.

Probes and Effectors are analogous to Sensors and Actuators, respectively. Indeed, Probes could be regarded as a type of Sensor (or vice versa) and Effectors could be regarded as a type of Actuator. We separate them so their roles are clear: Probes and Effectors interface between the managing and managed system whereas Actuators and Sensor exist within the managed system, between Controllers and Processes.

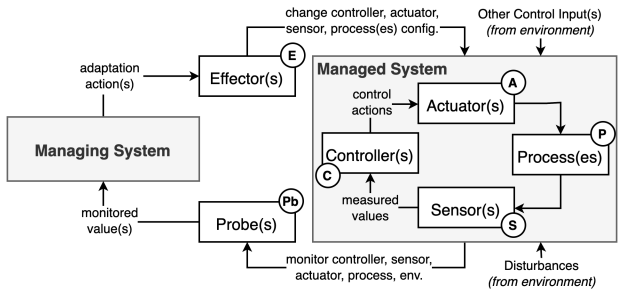

*5) Refine Managing System:* In the fifth step the managing system is refined to include Monitor, Analyzer, Planner, and Executor components. From a control-theoretic perspective, each of these components is regarded as a type of Controller that provides control actions to other elements. This step permits multiple Monitors, Analyzers, Planners, or Executors, depending on the specifics of the system being modelled. In the figure below, the details of the managed system are 'collapsed' to save space into a single block.

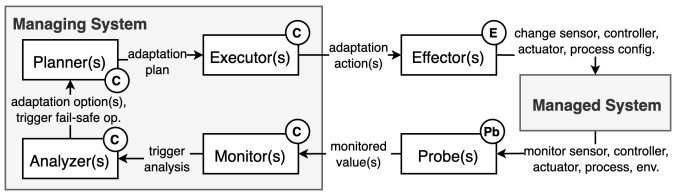

*6) Add Knowledge:* The Knowledge repository is an essential aspect of the MAPE-K reference model used within the managing system. In the sixth step the Knowledge repository is included in the control structure. The result is shown in Fig.3 below. The relationships between the MAPE components and the Knowledge repository are based on Weyns' description of a MAPE-K [1]. In a generic STAMP control structure, a Controller optionally includes a 'process model' that describes the Controller's internal representation of the Process(es) under its control. The Knowledge repository in the managing system has a similar purpose and is therefore regarded as a shared process model. Weyns describes four types of models that may appear within a Knowledge repository: a managed system model, an environment model, a goal model, and working models [1]. The managed system model represents the state of real managed system. The environment model describes the state of the environment. The (adaptation) goal model describes the objectives of the SAS and is used to generate and analyze adaptation options. Finally, the working models are used to perform specific analysis tasks, such as predicting the effect of an adaptation option on the managed system. Even if they are not explicitly included for a real system, their inclusion in this reference control structure serves as a useful reminder of the data that may be stored in the Knowledge repository.

*7) Include Higher-Order Adaptation:* It is possible that other control elements will influence the managing system, perhaps by changing the adaptation goals [1]. Such 'higher-order' adaptations might be of interest for SASs. Indeed, Leveson and Stringfellow both use higher-order controllers to describe human or technical decision making that influences the behaviour of the 'primary' controller [14], [48]. In this last step these higher-order control elements may be optionally included with connections to the managing system. The final reference model is shown in Fig.3.

### C. Analysis Obligations by Adaptation Type

Once a STAMP control structure has been modelled and the adaptation type has been determined, the STPA analysis proceeds to identify UCAs and loss scenarios. Depending on the type of the system, the 'obligations' described in this section guide the analysis. The obligations are described in the narrative below and summarized as causal factors in Fig.4. For brevity, the obligations are focused on adaptation, the managing system, and the relationship(s) between the managed and managing systems; other (non-adaptation) analysis obligations might be applicable.

*1) Type 0 Adaptations:* By definition, Type 0 adaptations do not interfere with the safety-related functions of the system, so it is unlikely that their nominal behaviour will cause a hazard occurrence. However, when identifying loss scenarios, analysts should consider the possibility that inadequate or incorrect operation of Effectors or Probes interferes with the managed system. For example, suppose a Probe makes too many network requests to measure attributes of a Controller such that critical tasks in the Controller are starved of processor time.

*2) Type I Adaptations:* Adaptation options for Type I adaptations are nominally chosen from a pre-determined set of safe options. However, the managing system might, due to a defect, provide an invalid (not from the known safe set) adaptation option. Therefore, when identifying UCAs the analyst should consider the possibility that Analyzer, Planner, or Executor, generates, selects, or applies an adaptation option that is not in the pre-determined set of safe adaptation options.

When identifying loss scenarios analysts should consider that an Effector might incorrectly apply an adaptation action such that an unsafe configuration produced. Additionally, since Probes might measure safety-related attributes of the managed system, the analysis should consider the possibility that a Probe interferes with a safety-related function of the managed system. Finally, the analysis should consider that the set of adaptation options in the Knowledge differ from the predetermined safe set of options, e.g., due to data corruption.

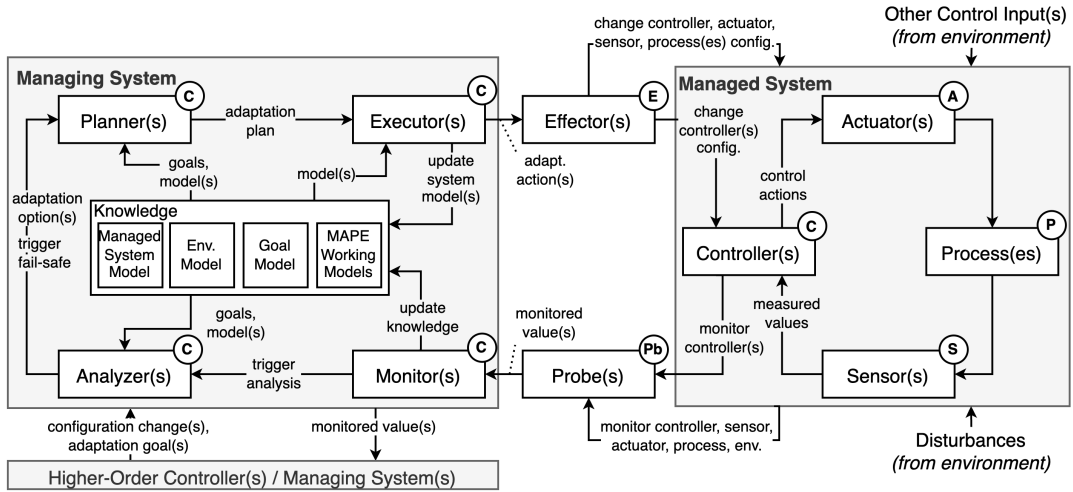

Fig. 3. STAMP reference model for self-adaptive systems.

*3) Type II Adaptations:* Type II adaptations are shown (at design-time) to be safe, subject to a set of constraints on the managed system or environment. If the constraints are satisfied, then the managed system is assured. Identification of UCAs should consider the possibility that the Analyzer, Planner, or Executor applies an unsafe adaptation option, including the possibility that an adaptation option is applied when its safety constraints are violated. Additionally, since selecting a safe adaptation option depends on the state of the operational environment or managed system, the possibility that the Monitor incorrectly or inaccurately updates, is delayed in updating, or misses updates to the Knowledge or unnecessarily triggers analysis should be considered. For example, if the Monitor misses updating the Knowledge about the environment, the Analyzer might falsely assume that the certain environmental constraints are satisfied and select an unsafe adaptation option. Finally, the choice of the next (safe) adaptation option might depend on the current system configuration, so the analysis should consider the possibility that an Executor misses, is delayed, or incorrectly updates the system model when applying an adaptation action.

Like for Type I, when identifying loss scenarios for Type II adaptations the analysis should consider the role that inadequate or incorrect operation of an Effector might have on the managed system. Additionally, the analysis should consider that inadequate operation might result in a Probe providing delayed, inaccurate, incorrect, or no data to the managing system. Finally, models of in the Knowledge might be incomplete, incorrect, or inconsistent.

*4) Type III Adaptations:* The analysis obligations for Type III adaptations are similar to those for Type II, but with the following additions. When identifying unsafe control actions, the analysis should consider the possibility that the Analyzer is delayed or does not trigger a fail-safe control safe action and the Planner or Executor are delayed in providing or do not provide the corresponding fail-safe system configuration. Similarly, when identifying loss scenarios, the analysis should consider the possibility that the Effector(s) are delayed or do not apply the fail-safe configuration changes. Finally, the analysis should consider the possibility that adaptation goals in the Knowledge are out-dated or in conflict such that unsafe adaptation options are provided by the Analyzer.

IV. EVALUATION

Evaluation of the STPA for SASs method is considered from three perspectives: applicability, repeatability, and effectiveness. Since this paper is the first to apply STPA to SASs, the evaluation focuses on applicability and offers preliminary considerations for repeatability and effectiveness.

*A. Applicability*

To evaluate applicability, we applied the method to two exemplar SASs and our own toy water heater system. Aside from the water heater [15], the exemplars were selected from a public list based on their prominence in the SAS community, the safety-related nature of the application, use of the MAPE-K reference model for the managing system, and availability of published descriptions of the system. The chosen systems were: Delta IoT wireless sensor network [1], [7], [50] and the UNDERSEA Unmanned Underwater Vehicle (UUV) [10], [18], [51]. The results of applying the method to the example systems are summarized in Table 1 which shows the adaptation type and the number of hazards, UCAs, loss scenarios, and mitigations identified for each system. Below we recount the application of the method to these systems and then provide additional remarks about applicability.

*1) STPA for Water Heater:* There is one hazard for the water heater system: *H1 - excessive temperature*. Since the water heater uses a basic feedback controller and MAPE-K architecture, the control structure for the water heater is nearly identical to the reference model in Fig.1, so for brevity it is not repeated here. The water heater's adaptation is Type II because the managing system picks from a pre-determined set of a controller gains, whose safety is dependent on current environmental conditions. Eight UCAs were identified. For example: *the Monitor is delayed in updating input flow knowledge such that the Analyzer is unaware of the latest data when analyzing env. conditions*, which could lead to an over-temperature event if the wrong conditions are used to pick the next set of controller gains. Also, 12 loss scenarios and 15 mitigations were identified.

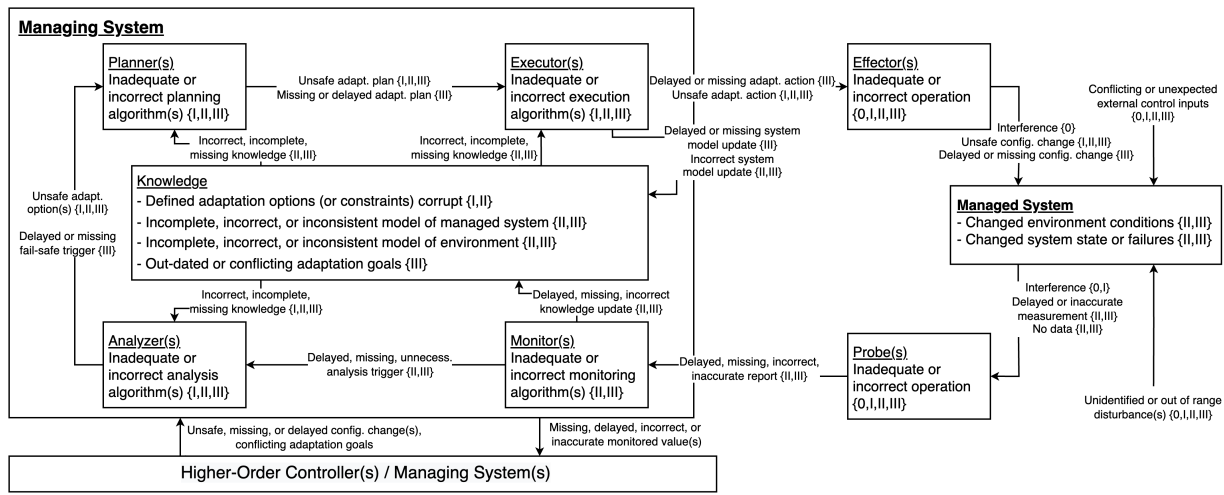

Fig. 4. Summary of causal factors for consideration when identifying loss scenarios; adaptation types shown in {...} braces.

TABLE I
SUMMARY OF ANALYSIS RESULTS FOR EACH SYSTEM.

| System | Type | Hzd. | UCA | Scen. | Mit. |
|---|---|---|---|---|---|
| Water Heater [15] | II | 1 | 8 | 12 | 15 |
| Delta IoT [7] | III | 2 | 29 | 33 | 48 |
| UNDERSEA UUV [10] | III | 3 | 45 | 60 | 50 |

*2) STPA for Delta IoT:* We identified two hazards for the Delta IoT system: *H1 - Missing Report to Security Personnel* and *H2 - Inaccurate Report to Security Personnel*. Using the modelling steps from Section III-B we prepared a control structure for Delta IoT (see top of Fig. 5). Rather than modelling all possible motes in the network, we used singular subject Mote that reads data from sensors in the facilities being monitored and propagates data and configurations from/to Child Motes. We classified the adaptations performed by Delta IoT as Type III because the adaptations change safety-related properties (e.g., transmission properties) and the system does not have a pre-determined set of configurations to pick from. 29 UCAs were identified for the system; for example: *the Planner does not provide an adaptation plan when the network performance has unacceptably degraded or a mote has failed*, which leads to hazard H1 occurring. Using the analysis obligations for a Type III system (see Fig. 4), we identified 33 loss scenarios; for example:

> Interference degrades network performance so some motes cannot transmit data packets [changed environment]; the Planner receives adaptation options from the Analyzer; the adaptation goals in the Knowledge are conflicting [conflicting goals]; the Planner is unable to pick a configuration that satisfies all goals and does not output any plan to the Executor [missing plan]; the network configuration is not changed [missing action, missing config. change]; some motes cannot transmit to their parents [missing control action]; an incident occurs but Security Personnel are not notified [H1].

Where the causal factors from the reference model are shown in square braces [...]. We identified 48 mitigations, expressed as shall statements, that Delta IoT could implement to mitigate the identified UCAs and loss scenarios.

*3) STPA for UNDERSEA UUV:* We identified three hazards for the UNDERSEA UUV: *H1 - Loss of separation with terrain*, *H2 - Missing data*, and *H3 - Inaccurate data*. Applying the modelling steps from Section III-B produced the control structure (see bottom of Fig. 5). The adaptations performed by UNDERSEA are Type III because they impact safety-related aspects of the system and the adaptation options are generated and verified (using probabilistic model checking) at run-time. 45 UCAs were identified; for example: *the Analyzer provides candidate configurations with the UUV speed set too high for the sensors when one of the sensor's acquisition rates has changed*, which could lead hazard H3 occurring. Using the obligations for a Type III system, 60 loss scenarios were identified, for example:

> A sensor's rate degrades and it is reported to the Monitor by the Probe; but the Knowledge's model of sensor accuracy v. speed is inaccurate [incorrect knowledge]; the Analyzer uses this model to pick a sensor configuration and speed that is incompatible [unsafe adapt. option]; the new configuration is applied and the UUV Motion Controller increases the speed [inappropriate control action]; the sen-sors' accuracy degrades as the speed increases and inaccurate measurements are recorded [H3].

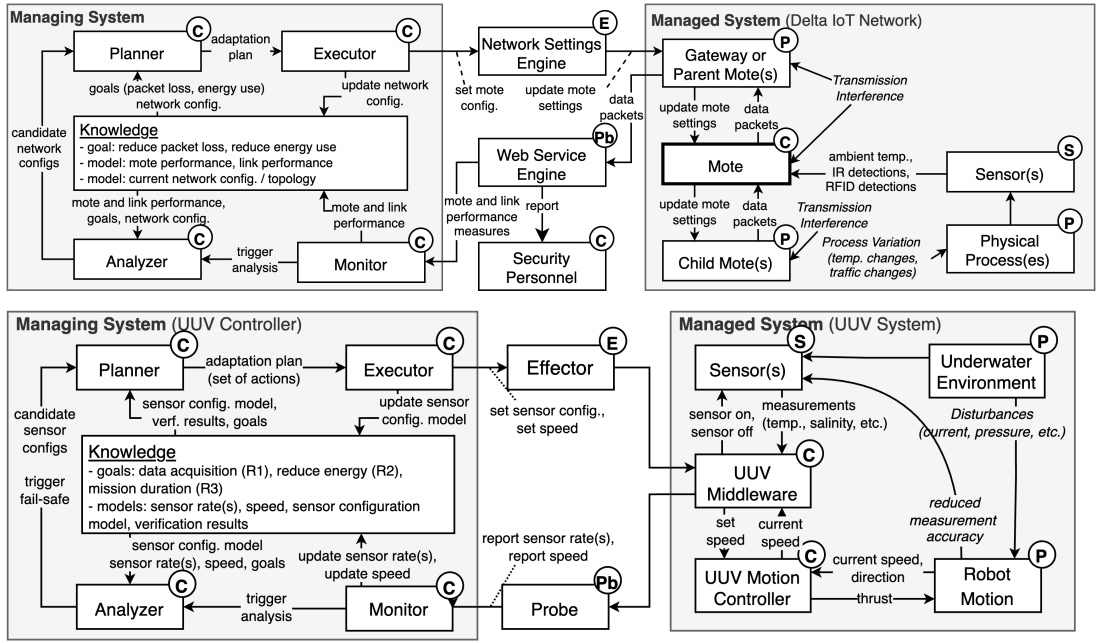

Fig. 5. Control structures for exemplar systems; top: Delta IoT [7], bottom: UNDERSEA UUV [10].

In total, 50 mitigations were identified that UNDERSEA could implement to address the UCAs and loss scenarios.

*4) Observations for Applicability:* We offer three observations about the applicability of the method. First, applying the modelling steps and STAMP reference model in Section III-B produced control structures that had a useful level of detail. The control structures were not trivial, which could lead to missed UCAs or loss scenarios, and were also not unmanageably complex, which could lead to UCAs or loss scenarios that are not credible or difficult to understand. Moreover, we did not encounter an components, functions, or behaviours of the systems (as described by the original authors) that could not be captured in the control structure at our chosen level of abstraction.

Second, the number of UCAs identified are comparable to other published accounts of applying STPA to (non-SAS) systems [42], [52]–[55]; though the number is slightly lower than if the analysis was performed on a real-world system. This is due to our analyses focusing on self-adaptive aspects (MAPE-K, etc.) of the system rather than the details of the managed system. For loss scenarios, since there are combinatorially many, the analyses opted to identify only one or two scenarios that contextualize the UCAs and demonstrate that the method produces credible loss scenarios.

Third, the examples covered a range of system and adaptation types. The examples include a wireless sensor network, an unmanned robotics system, and a basic feedback controller. This suggests that the method is applicable to a breadth of SASs. Moreover, the adaptation types considered were II and III, which cover (for the most part) the adaptation obligations for types 0 and I as well.

### B. Repeatability

For hazard analysis methods, repeatability refers to whether different analysts will arrive at the same results when applying the same method to the same system [56]. One way to assess repeatability is to perform studies where multiple analysts independently apply the method and then compare the results, e.g., are similar UCAs and loss scenarios identified? Such a study is beyond the scope of this paper. Instead, we offer two observations regarding repeatability.

First, control structure modelling is subject to variability arising from the perspectives of analysts. Since the control structure is the basis of STPA, it has a significant impact on the outcome. The derivation steps in Section III-B were originally intended to serve as a proof-by-construction for our reference control structure. However, during the evaluation we found the derivation steps helpful to guide modelling. Specifically, the steps helped to organize our thinking about the relationship(s) between the managed and managing system and often prompted revisits the source materials to confirm details. This unexpected, but not unwelcome, observation suggests that the modelling steps enhance the repeatability of STPA when applied to SASs that use the MAPE-K reference model.

Second, the set of analysis obligations in Section III-C aid in repeatability by ensuring that a minimum set of UCAs and causal factors are considered. During the evaluation, we found the obligations helped to guide our thinking and prompted us to re-consider SAS-specific aspects that might not have been obvious without them.

*C. Effectiveness*

In general, it is difficult to assess whether a hazard analysis method is effective at finding all causes of a hazard. Longitudinal or retrospective studies that compare the causes identified by analysis methods to hazard occurrences reported over the lifetime of a system are one way to assess effectiveness. However, such studies are practically difficult to conduct. Instead, researchers often compare *between* methods to determine which one produces more meaningful results. Regarding STPA, several authors have concluded that it is an effective hazard analysis method, especially when used to assess interactions between components at the system level [37], [38], [52], [54], [57]–[59]. Since the STPA for SASs extends STPA it is reasonable to believe that it be at least as effective the original STPA method. However, this depends on the assumption that the extended method has not departed from original STPA in a manner that compromises effectiveness, i.e., *are we still doing STPA?* We address this question in two parts. First, the extended method includes all the steps of original STPA. Though the extension adds a step (Classify Adaptation(s)) and reference inputs (control structure, analysis obligations), it has not removed or altered the original analysis steps. Moreover, the application of the extended method to examples faithfully applied the original STPA steps and produced UCAs and loss scenarios as wanted. Second, the extended method does not prevent an analyst from applying STPA as originally intended. The STAMP reference model derived in Section III-B employs Leveson's modelling constructs (Processes, Controllers, Actuators, Sensors) and does not preclude modelling sophisticated control structures that occur in real-world systems. The obligations in Section III-C describe factors that must be *minimally* considered for a SAS, but this not prevent consideration other aspects of the system.

A question about effectiveness remains: *is STPA for SASs more effective than the original STPA method for analyzing SASs?* Answering this question is beyond the scope of this paper. One approach would be to perform a 'control trial'. A control group would apply original STPA and an intervention group would apply the method as described in this paper. Results could be compared based on the number of UCAs and loss scenarios, complexity of control structures, and analysis effort.

## V. Related Work

First, this section reviews existing hazard analysis approaches for SASs. Then, the existing approaches are compared to the STPA for SAS method introduced in this paper.

Some authors have described design-time only methods for SAS hazard analysis. Witte *et al.* combine fault trees, attack trees, dataflow models, and deployment models to analyze safety and security events in a SASs' managed system at design-time [28]. Their fault and attack trees are created by experts or from vulnerability databases. They use graph transformation rules to describe the effect of adaptations on the model. Model checking over the rules checks whether the managed system can enter a potentially dangerous state as a result of adaptation. Priesterjahn *et al.* assess, at design-time, whether a self-healing system can respond to a failure before it contributes to the occurrence of a hazard using Timed-Failure Propagation Graphs (TFPGs) [26]. First, they model the managed system using UML deployment diagrams; then minimum cutsets are derived from an manually created fault tree; for each cutset they generate a TFPG and analyze its timing properties. Alder *et al.* use component fault trees, markov chains, and MARS adaptation models to compute, at design-time, the probability that the managed system is in a hazardous configuration [25]. Finally, Wotawa discusses the use of FMEA for design-time analysis of self-repairing systems [27]. Like STPA for SASs, the above methods are applicable exclusively at design-time.

Other authors have developed 'hybrid' methods that have both design- and run-time phases. Aslanesfat *et al.* use design-time fault tree and markov process models that are augmented at run-time with sensor data to estimate the probability of failure of the managed system [31]. Brunner *et al.* consider re-configurable ("plug and fly") avionics systems and propose an engineering process that spans both design- and run-time [30]. At design-time they use FTA or FMEA to identify hazard causes for each component which are then incorporated into run-time analysis based on the current system configuration. Dobaj *et al.* created INSpIRA, a method for the analysis of safety and security properties [29], [60]. INSpIRA uses design-time FMEA to find cause-effect relationships which are provided to a run-time 'execution engine' that executes tests on the managed system to determine if a failure mode has occurred. Bhardwaj and Liggesmeyer used a design-time HAZOP to analyze component '"deviations" that are then stored in the system's Knowledge and used at run-time to assess the risk associated with new architectural configurations of the managed system [61]. Schneider and Trapp introduce conditional safety certificates (ConSerts) that can be dynamically composed at run-time in a ConSert Tree (like a fault tree) to determine if the current selection of components in the managed system collectively achieve an overall system integrity goal [17], [62]. The ENTRUST method uses design- and run-time system models of the managed and managing system to verify adaptations via model-checking and simulation [18]. At design-time, models for the managing system are verified while at run-time models for the managed system (accounting for the current configuration) are verified. The evidence generated from verification is incorporated in to a dynamic assurance case for the system. Though ENTRUST is not strictly a hazard analysis method (in the sense that it produces a list of hazard causes), it has many similarities to existing methods and aims to achieve the same overall goal of assuring a SAS.

All the hazard analysis analysis approaches include a design-time analysis that uses established methods such as FMEA, FTA, or HAZOP. Such methods employ a limited notion of accident causality that is focused on component failure [32]. By contrast, STPA for SASs uses STAMP's wholistic model of accident causality that accounts for non-failure hazard causes, such as component interactions [14].

Additionally, the existing hazard analysis approaches (excluding ENTRUST), focus on hazard causes arising from the managed system and assume that adaptations are faithfully carried out by the managing system. Also, they do not consider SAS-specific aspects when generating hazard causes. STPA for SAS overcomes this limitation by including the managing system, and the adaptations it performs, in the scope of the analysis via adaptation classification, reference models, and analysis obligations.

## VI. Discussion

This paper has described an extension to STPA for SASs. The extension includes a reference model and derivation procedure that can be applied to arrive at a control structure for a SAS and a set of analysis obligations to be considered by analysts. The extension overcomes limitations of existing hazard analysis approaches arising for their model of accident causality and focus on the managed system. Below the method's industrial significance, limitations, and threats to validity are discussed.

### A. Industrial Significance

In addition to its theoretical contribution, STPA for SASs has practical significance. The following remarks are based on our industrial experience with conventional systems in health, rail, automotive, energy, and robotics.

First, developers of safety-critical systems must perform hazard analyses to achieve compliance with technical standards. It is necessary to apply methods 'by the book' to manage risk. Though STPA is recognized as an effective method, there is no guidance for practitioners on how to apply it to SASs, i.e., there is no 'book'. The contributions in this paper can guide practitioners. In our experience, the availability of such resources reduces cost and improves the quality of hazard analyses.

Second, *completeness* is a common concern encountered when performing hazard analysis, i.e., *have we thought of everything?* Guide phrases are often used to prompt thinking about hazard causes (e.g., the phrase "stuck open" is often applied to valves). STPA uses guide phrases when identifying UCAs and loss scenarios such as "too late" and "missing control action" [14]. However, these phrases are not SAS-specific. Our analysis obligations provide SAS-specific considerations, which will help increase confidence in the analysis.

Finally, safety requirements are a cornerstone of the engineering process prescribed by industrial standards [20], [22], [63]. To achieve compliance, requirements must be specified and verified for the managed and managing systems. By applying STPA to SASs, practitioners can elicit these requirements.

### B. Limitations and Threats to Validity

The proposed method and the evaluation is subject to limitations and threats to validity, these are discussed below.

*1) Design-Time Method:* Currently, STPA for SASs is a design-time method and has not been used at run-time. Despite this limitation, we would like to point out the importance of design-time analysis [64]. First, from a practical perspective, deploying a safety-critical SAS with assurances based only on run-time hazard analysis might be a challenging proposition for industrial stakeholders (regulators, the public, etc.) since established standards demand design-time analysis. Second, design-time analysis can be used to inform subsequent engineering activities; for instance, allocating more effort to critical components. Third, existing hybrid methods depend on design-time analysis to establish a baseline that is used at run-time, which necessitates design-time methods. Fourth, from a research perspective, new design-time methods might lead to insights for run-time hazard analysis. Finally, not all SASs require run-time analysis. When adaptations are well-defined, design-time methods that consider the managed and managing system (like STPA for SAS) might be adequate.

*2) MAPE-K Architecture:* The proposed method is dependent on the managing system using the MAPE-K reference model. While this is a popular model, some SASs might use other approaches. The proposed method might not be applicable to non-MAPE-K designs.

*3) Repeatability:* Though we have argued that aspects of the method, such as the reference model and analysis obligations, improve repeatability, this has not been formally assessed. The possibility that different analysts will produce widely different results cannot be discounted.

*4) Effectiveness:* The effectiveness of the proposed method was not formally evaluated. It is possible that this extension to STPA does not improve outcomes (compared to just STPA on its own). Though there is a case based on the studies of STPA, since we have not performed a formal comparison with other SAS-specific methods, we cannot conclude that this method is superior to other published design-time analysis methods.

## VII. Conclusion

This paper has proposed a design-time method extension that tailors STPA to SASs that employ the MAPE-K reference model. The results of applying the method to three examples show that STPA is an applicable method for this class of SASs. Additionally, preliminary evaluation suggests that the proposed method is both repeatable and effective, though further studies are required to confirm this is the case.

Future work might focus in the following areas: 1) using the framework described in this paper to perform run-time hazard analysis; 2) broadening the reference models to include other architectures beyond MAPE-K; and 3) performing more rigorous studies of the method's repeatability and effectiveness.